\documentclass[letterpaper, 10pt, conference]{IEEEconf}      

\IEEEoverridecommandlockouts                              
\usepackage[usenames,dvipsnames]{xcolor}
\usepackage{graphicx} 
\usepackage{mathptmx} 
\usepackage{times} 
\usepackage{amsmath} 
\usepackage{amssymb}  
\usepackage{pstricks,pst-plot,psfrag}
\usepackage{units}
\usepackage{multirow}
\usepackage[draft]{hyperref}
\usepackage[]{mcode}
\usepackage{booktabs}
\usepackage{balance}
\usepackage[acronym]{glossaries}
\usepackage{cite}
\usepackage{lettrine}
\usepackage{import}
\usepackage{tikz}
\usetikzlibrary{positioning}
\usetikzlibrary{calc}

\definecolor{lightblue}{rgb}{0.60784,0.76078,0.90196}
\definecolor{darkblue}{rgb}{0.26667,0.44706,0.76863}
\definecolor{lightgreen}{rgb}{0.66275,0.81569,0.55686}
\definecolor{darkgreen}{rgb}{0.43922,0.67843,0.27843}
\definecolor{orange}{rgb}{0.92941,0.49020,0.19216}
\definecolor{yellow}{rgb}{1.00000,0.75294,0.00000}
\definecolor{grey}{rgb}{0.64706,0.64706,0.64706}
\definecolor{purple}{rgb}{0.51373,0.23529,0.04706}
\newacronym{abk:amod}{AMoD}{Autonomous Mobility-on-Demand}
\newacronym{abk:iamod}{I-AMoD}{intermodal \gls{abk:amod}}
\newacronym{abk:bpr}{BPR}{the Bureau of Public Roads}
\newacronym{abk:ca}{CA}{congestion-aware}
\newacronym{abk:cara}{CARS}{congestion-aware routing scheme}
\newacronym{abk:cepamods}{CEPAMoDS}{Convolutional Energy Predicting AMoD Scheduler}
\newacronym{abk:cs}{CS}{charging station}
\newacronym{abk:ffcs}{FFCS}{free floating car sharing systems}
\newacronym{abk:ghg}{GHG}{greenhouse gas}
\newacronym{abk:kpi}{KPIs}{Key Performance Indicators}
\newacronym{abk:mcfp}{MCFP}{multi-commodity flow problem}
\newacronym{abk:milp}{MILP}{mixed-integer linear program}
\newacronym{abk:spp}{SPP}{shortest path problem}
\newacronym{abk:kdspp}{k-dSPP}{k-disjoint \gls{abk:spp}}
\newacronym{abk:soc}{SoC}{state of charge}
\newacronym{abk:vrp}{VRP}{vehicle routing problem}
\newacronym{abk:v2g}{V2G}{vehicle-to-grid}

\newcommand{\Pru}{P_\mathrm{rush}}
\newcommand{\Pch}{P_\mathrm{chill}}
\newcommand{\flexbrac}[1]{\if\relax\detokenize{#1}\relax \else (#1) \fi}
\newcommand{\flexcomma}[1]{\if\relax\detokenize{#1}\relax \else ,#1 \fi}

\newcommand{\vP}{\mathbf P}

\newcommand{\cY}{\mathcal{Y}}

\newcommand{\sN}{\mathbb{N}}

\newcommand{\sR}{\mathbb{R}}

\newcommand{\bx}{\bar{x}}

\newcommand{\bd}{\bar{d}}
\newcommand{\bs}{\bar{s}}

\newcommand{\by}{\bar{y}}
\newcommand{\cbY}{\bar{\mathcal{Y}}}

\newcommand{\bK}{\bar{K}}

\newcommand{\one}{\mathbb{1}}
\newcommand{\zero}{\mathbb{0}}
\newcommand{\Pho}{P_\mathrm{home}}
\newcommand{\Pgo}{P_\mathrm{go}}
\newcommand{\Pe}{\vP^\mathrm{e}}
\newcommand{\kref}{k_\mathrm{ref}}
\newcommand{\kinf}{k_\mathrm{inf}}
\newcommand{\kpoor}{k_\mathrm{poor}}
\newcommand{\krich}{k_\mathrm{rich}}
\newcommand{\kwealthy}{k_\mathrm{wealthy}}

\DeclareMathOperator*{\argmin}{arg\,min}
\newcommand{\summe}[2]{\sum_{\scriptstyle\mathclap{#1}}^{\scriptstyle\mathclap{#2}}}
\usepackage{mathtools}
\usepackage{dsfont}
\usepackage{booktabs}
\usepackage{bbm}
\usepackage{bbold}
\usepackage{cuted}
\usepackage{comment}
\let\proof\relax
\let\endproof\relax

\usepackage{amsthm}
\newtheorem{assump}{Assumption}
\newtheorem{theorem}{Theorem}[section]
\newtheorem{definition}[theorem]{Definition}
\newtheorem{problem}{Problem}
\newtheorem{lemma}[theorem]{Lemma}

\makeatletter
\newcommand{\pushright}[1]{\ifmeasuring@#1\else\omit\hfill$\displaystyle#1$\fi\ignorespaces}
\newcommand{\pushleft}[1]{\ifmeasuring@#1\else\omit$\displaystyle#1$\hfill\fi\ignorespaces}
\makeatother
\newif\ifmargincomments 
\margincommentstrue

\newif\ifextendedversion 
\extendedversiontrue

\ifmargincomments

\else

\fi

\title{\LARGE \bf
Urgency-aware Optimal Routing in Repeated Games\\ through Artificial Currencies
}

\author{Mauro Salazar$^{1}$, Dario Paccagnan$^{2}$, Andrea Agazzi$^3$, W.P.M.H. (Maurice) Heemels$^{1}$
\thanks{$^1$Control Systems Technology group, Department of Mechanical Engineering,
Eindhoven University of Technology, The Netherlands, e-mail:
{\tt\small \{m.r.u.salazar,w.p.m.h.heemels\}@tue.nl}.}
\thanks{$^2$Computational Optimization group, Department of Computing, Imperial College London, United Kingdom, e-mail: {\tt\small d.paccagnan@imperial.ac.uk}.}
\thanks{$^3$Department of Mathematics, Duke University, United States, e-mail: {\tt\small andrea.agazzi@duke.edu}.}
}

\begin{document}
\maketitle

\begin{abstract} 
When people choose routes minimizing their individual delay, the aggregate congestion can be much higher compared to that experienced by a centrally-imposed routing. Yet centralized routing is incompatible with the presence of self-interested agents. How can we reconcile the two?
In this paper we address this question within a \emph{repeated game} framework and propose a fair incentive mechanism based on artificial currencies that routes selfish agents in a system-optimal fashion, while accounting for their temporal preferences.
We instantiate the framework in a parallel-network whereby agents commute repeatedly (e.g., daily) from a common start node to the end node.
Thereafter, we focus on the specific two-arcs case whereby, based on an artificial currency, the agents are charged when traveling on the first, fast arc, whilst they are rewarded when traveling on the second, slower arc.
We assume the agents to be rational and model their choices through a game where each agent aims at minimizing a combination of today's discomfort, weighted by their urgency, and the average discomfort encountered for the rest of the period (e.g., a week).
We show that, if prices of artificial currencies are judiciously chosen, the routing pattern converges to a system-optimal solution, while accommodating the agents' urgency.
We complement our study through numerical simulations.
Our results show that it is possible to achieve a system-optimal solution whilst reducing the agents' perceived discomfort by 14-20\% when compared to a centralized optimal but urgency-unaware policy.
\end{abstract}

\section{Introduction}
\lettrine{M}{obility} systems are currently facing significant challenges due to users’ dissatisfaction, traffic congestion and environmental pollution~\cite{TuttleCowles2014,EPA2018}. At the same time, the advent of  driving technologies, the internet of things, the concept of sharing economies and new automotive technologies is leading to structural transformations in the way we conceive mobility, providing us with unprecedented opportunities to handle the aforementioned challenges. For instance, intermodal autonomous mobility-on-demand systems---fleets of robotaxis servicing travel demand jointly with public transit---are a promising solution for urban scenarios, as they combine the high-efficiency of public transportation systems with the point-to-point mobility service provided by fleets of connected autonomous vehicles.
Whilst the possibility of routing customers through socially-optimal intermodal routes can significantly improve transportation efficiency~\cite{SalazarLanzettiEtAl2019}, this requires users to sacrifice their individual welfare for the ``greater good''~\cite{Wollenstein-BetechHoushmandEtAl2020}. The central issue 
revolves around the inherent tension between the drivers' individual objective (e.g., minimizing the travel-time from A to B) and the societal goal (e.g., minimizing the city-wide congestion). In this respect, it is well known  that traffic patterns arising from self-interested decision making are often inefficient~\cite{RoughgardenTardos2002}, and this is, to a high degree, what we experience in every densely populated city.

\begin{figure}[t]
	\includegraphics[width=\columnwidth]{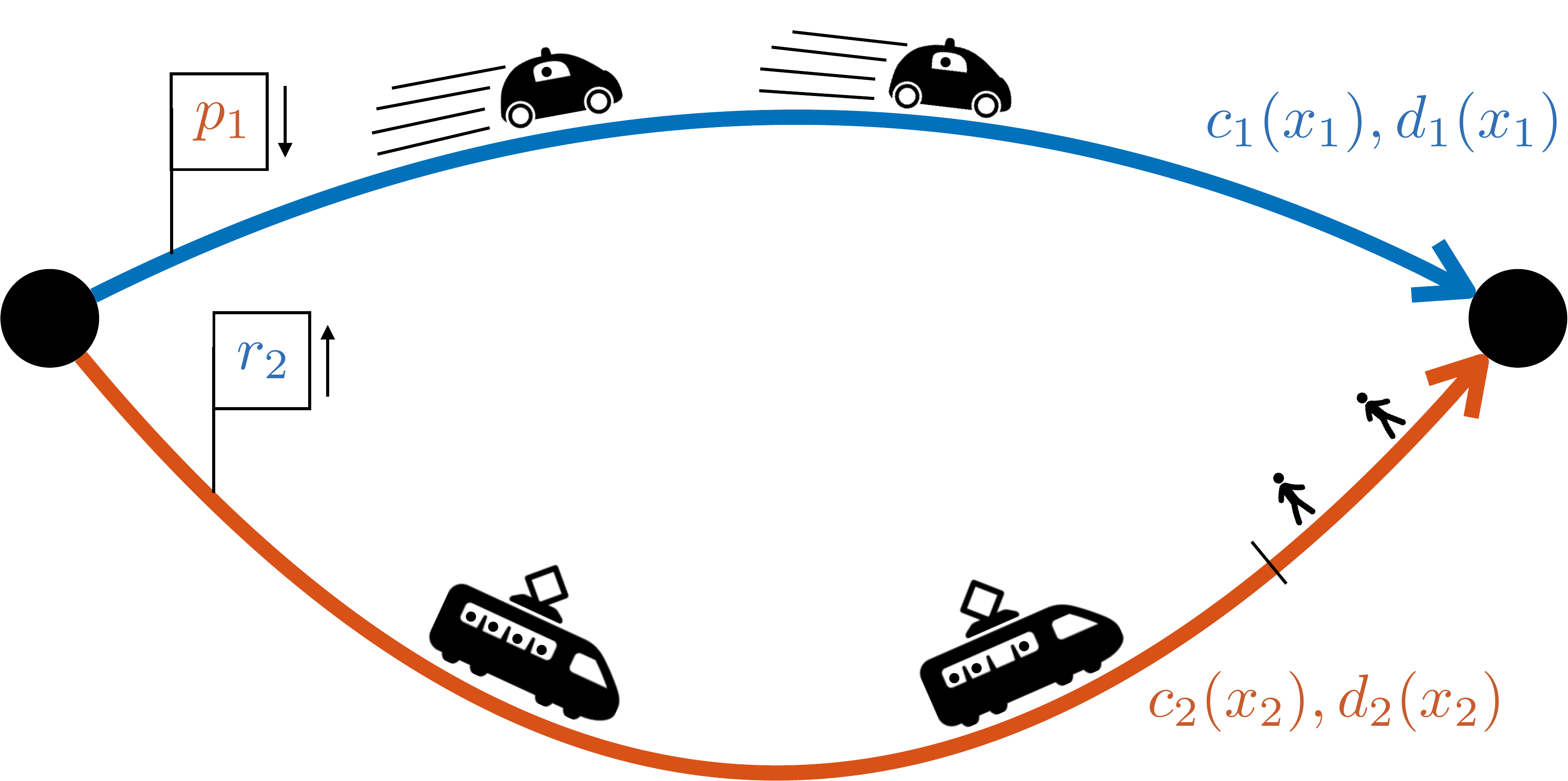}
	\caption{
		Network with one origin and one destination node connected by two arcs. Each arc $j$ is traversed by $x_j$ people per unit time, resulting in a societal cost $c_j(x_j)x_j$ and a discomfort cost $d_j(x_j)$ perceived by the agents. Each arc is assigned a price $p_j$. In this particular case, the agents pay $p_1>0$ to take the first, fast route, and receive $r_2=-p_2>0$ when traveling the second, slower route.
	}
\label{fig:ODwithMultiarcs}
\end{figure}

A promising solution to alleviate these issues is to employ monetary tolls~\cite{Pigou1920, BergendorffHearnEtAl1997, Morrison1986, BrownMarden2017}. However, approaches based on congestion pricing are associated with two fundamental drawbacks: i) they discriminate users with lower incomes; ii) they do not account for individual preferences such as the users' temporary urgency and needs. Against this backdrop, this paper presents an incentive scheme based on an artificial currency---here called \emph{Karma}, borrowing the terminology from~\cite{CensiBolognaniEtAl2019}---to align the routing of self-interested users with the system-optimum allocation, whilst accounting for their temporal individual needs.
Specifically, our framework is based on a currency that can neither be bought nor exchanged, but only spent or gained when traveling.
On the microscopic level, it gives each user equal possibilities to choose when to be self-interested---selecting the fastest path for a toll (e.g., the upper one in Fig.~\ref{fig:ODwithMultiarcs})---and when to be altruistic---traveling the slower path for a reward (e.g., the lower one in Fig.~\ref{fig:ODwithMultiarcs})---whilst, on the mesoscopic level, it aligns the average population behavior to the system optimum.

\emph{Related work:}
The interest in the design of tolls dates back to the work of Pigou~\cite{Pigou1920}. Since then, a large body of literature pertaining to the fields of transportation and economics has developed this approach~\cite{Pigou1920, BergendorffHearnEtAl1997, Morrison1986}. When considering the continuous flow approximation of the classical congestion game model \cite{Rosenthal1973} (typically employed to describe mesoscopic traffic patterns), the marginal cost mechanism~\cite{BeckmannMcGuireEtAl1955} produces local tolls ensuring that every equilibrium allocation coincides with the system optimum. The problem of optimal tolling is much more challenging when considering the original congestion game model. In this context, the recent result in~\cite{PaccagnanChandanEtAl2019} derives optimal local tolls that minimize the system inefficiency and can be considered the counterpart of marginal cost tolls for discrete congestion games.
Designing tolling mechanisms that account for the agents' sensitivity distribution is particularly challenging even if considering congestion games with a continuum of agents, unless the exact sensitivity of each agent is fully known to the designer.
In this respect, the results are limited to providing efficiency bounds for existing tolling mechanisms~\cite{MeirParkes2016}, to deriving optimal tolls when the sensitivity distribution is a piecewise constant function, or to proving the existence of optimal tolls for general distributions~\cite{FleischerJainEtAl2004}. Our work significantly departs from all these studies for two fundamental reasons: First, we utilize an artificial currency that can neither be bought nor exchanged; second, we account for the agents' sensitivity in a repeated game scenario with the objective of achieving the system optimum in the long run.

Whilst the use of artificial currency as a means to coordinate self-interested decision makers has recently attracted significant attention, see e.g., \cite{Prendergast2016, CensiBolognaniEtAl2019, GorokhBanerjeeEtAl2019}, most of the existing works are based on auction mechanisms and have not been applied to managing traffic routing. Perhaps closest in spirit to our work is~\cite{CensiBolognaniEtAl2019}, whereby the authors introduce an artificial currency called Karma to coordinate the behavior of competitive decision makers and allocate shared resources (e.g., intersection coordination).
Their central idea consists in allowing for agents to pass on using resources today, in exchange for Karma, which will allow them to claim the resource in the future. While our work is based on a similar philosophy, it departs significantly from the former in that we envision its application to mobility systems (e.g., traffic routing). As such and in contrast to~\cite{CensiBolognaniEtAl2019}, each resource is associated to a measure of its quality (e.g., travel time), and our model accounts for the presence of more than two decision makers.
Moreover, agents are not required to submit Karma bids, whilst their routing is controlled through simple payment transactions.

\emph{Statement of contributions:}
This paper introduces a repeated game framework and leverages an artificial currency to route agents in a system-optimal fashion while accounting for their temporal preferences.
In particular, we focus on repetitive events, e.g., daily commutes, whereby traveling agents choose between two possible routes, as shown in Fig.~\ref{fig:ODwithMultiarcs}, depending on their daily sensitivity (e.g., urgency) to the discomfort incurred when traveling (e.g., travel time).
Considering a central operator who wants to align the mesoscopic agents' behavior with the societal optimum, we first instantiate a simple pricing framework that is not relying on complex auction mechanisms, but gives the agents full freedom of choice.
Second, we propose a decision-model for rational agents for which we compute the best response strategy in closed form, demonstrating it to exhibit a clear structure.
Third, we show that the only possible static pricing policy resulting from Karma-conservation arguments makes the system optimum the only and globally asymptotically stable equilibrium of the system, in the sense that the population behavior will converge to it for any initial Karma allocation.
It is important to highlight how our mechanism does not require any information on the underlying sensitivity distribution, i.e., it is distribution-free.
Finally, we simulate our scheme in different case studies, validating our theoretical findings numerically.

\emph{Organization:}
The remainder of this paper is structured as follows: We instantiate our framework in Section~\ref{sec:genmodel}, where we introduce the repeated game problem on a parallel arcs network for a central operator and for an individual agent.
From Section~\ref{sec:twoarcs} onward we focus on a two-arcs parallel network, by first deriving the optimal static prices via necessary Karma-conservation conditions.
Section~\ref{sec:bestresponse} computes the best response strategy for the agent analytically, which is used in Section~\ref{sec:mesoscopic} to study the average aggregate behavior of a population of agents playing the best response strategy, showing that it converges to the desired optimal flows.
We validate our findings with numerical simulations for different case studies in Section~\ref{sec:results}, and conclude the paper in Section~\ref{sec:conclusion} with a discussion and an outlook on future research.

\section{Problem Formulation}\label{sec:genmodel}
This section introduces the routing problem from both the mesoscopic perspective of a central operator who wants to minimize the societal costs, and from the individual perspective of an agent aiming at minimizing the discomfort incurred when traveling under the proposed scheme.

Our framework combines three ingredients: i) a transportation network represented by a digraph, ii) cost functions representing the societal and personal costs (a measure of discomfort such as travel time) resulting from the aggregate route choices of the agents, and iii) a static pricing policy based on the artificial currency Karma.

\subsection{Central Operator's Problem}\label{subsec:parallel}
Consider a parallel road-network, consisting of a single origin and destination node but connected by $n\in\{1,2,\dots\}$ directed arcs, e.g., the example in Fig.~\ref{fig:ODwithMultiarcs}, whereby $n=2$.
Such a digraph can be used to describe a daily commute, but our framework lends itself to model more general resource-allocation problems whereby the quality of each resource depends on its usage-level.

From a mesoscopic perspective, the agents want to reach the destination from the origin traveling through one of the $n$ arcs in the digraph at each (discrete) time $t\in\sN$.
Thereby, the component $x_j(t)$ of vector $x(t)\in[0,1]^n$ represents the fraction of users crossing arc $j\in\{1,2,\dots,n\}$ at time $t$.
Every day, each agent has a probability $\Pho\in[0,1]$ not to travel.
This way, the probability for an agent to travel is $\Pgo = 1-\Pho$ and $\sum_{j=1}^n x_j(t)=\Pgo$ people will travel per day.
Crossing each arc $j$ entails a specific societal cost per person~$c_j(x_j(t))$ dependent on the number of people traversing it $x_j(t)$.
Moreover, it also causes each person a discomfort $d_j(x_j(t))$ (e.g., travel time).
We assume $c:[0,1]^n\to \sR_{+}^n$ and $d:[0,1]^n\to \sR_{+}^n$ to be monotonically increasing functions for each arc~$j$, where \mbox{$\sR_{+}=[0,\infty)$} is the set of real non-negative numbers.
In general, the societal cost $c(x)$ and the discomfort $d(x)$ need not be aligned---for instance, the former might represent energy consumption and the latter travel time.

We assume the presence of a central operator (e.g., a municipal authority) who needs to design incentives so that the aggregate flows converge to the minimizer of the total societal cost $c(x)^\top x$.
\begin{problem}[Central Operator's Problem]\label{prb:SO}
The central operator aims at routing customers so the aggregate route choice $x(t)$ converges to
{\begin{subequations}
\small
\begin{align}
x^\star \in \arg \min_{x\in[0,1]^n} \; & c(x)^\top x\\
\text{s.t. } & \one^\top x = \Pgo.
\end{align}
\end{subequations}}
\end{problem}

In order to steer peoples' behavior towards a social optimum $x^\star$, the central operator endows each agent with the artificial currency Karma and sets a price $p_j\in\sR$ to cross each arc $j$.
Agents are not permitted to buy or exchange Karma, and they can only select the arcs keeping their Karma-level non-negative.
Crucially, some of the arcs have negative prices, so that agents will be rewarded when crossing them.

From a microscopic perspective, we model the routing choice for an individual agent $i$ at time $t$ as $y^i(t)\in\{0,1\}^n$ being a vector with $y_j^i(t)=1$ if person $i$ decides to cross arc $j$ at $t$, and 0 otherwise.
Thereby, a non-traveling agent is modeled as $y^i(t)=\zero$.
Given a scenario with $M$ agents, their individual decisions are linked to the mesoscopic flows as $x(t) = \frac{1}{M}\sum_{i=1}^M y^i(t)$.
In particular, assuming that at time $t$ each agent $i$ owns an amount $k^i(t)\in\sR_+$ of Karma, given a routing choice $y^i(t)$, her Karma will be updated as $k^i(t+1) = k^i(t) - p^\top y^i(t)$.

\subsection{Individual Agent's Problem}
On the microscopic level, we assume individual agents to make choices in order to minimize their traveling discomfort without reaching a negative level of Karma.
In contrast to conventional monetary tolling schemes, the individual agent's problem cannot be captured in a static setting: In our framework, the agents are playing against their future selves, and must account for their future preferences to decide when to use Karma and when to gain it. Specifically, on some days they might be more sensitive to discomfort than on other days---for instance, when going to an important meeting, travel time may be more important than when simply commuting to the working place.
To this end, we define the sensitivity that a person $i$ might have at time $t$ w.r.t.\ discomfort as $s^i(t)\in\sR_+$ and use it as a weighting factor. This way, the cost perceived by agent $i$ when crossing arc $j$ is $s^i(t)\cdot d_j(x_j(t))$.
Hereby, we assume $s^i(t)$ to be i.i.d.\ extractions (w.r.t.\ $i$ and $t$) from an underlying common probability distribution with  probability  density function \mbox{$\rho:[s_\mathrm{min},s_\mathrm{max}]\to\sR_+$}, support set $[s_\mathrm{min},s_\mathrm{max}]\subseteq\sR_+$ and average value $\bs$.
For the sake of simplicity, from now on we drop dependence on $t$, $x$ and $i$ whenever it is clear from the context.

We assume that each agent will choose the route associated with the least discomfort perceived on the respective day of traveling and the one expected in the $T$ days to follow, whilst accounting for her Karma allowance.
Specifically, we model the decision of each agent in the present moment as $y^i\in\{0,1\}^n$, whilst we condense her future behavior into the fractional variable $\by^i\in[0,1]^n$.
We assume the agents to consider today's discomfort on the arcs $d(x(t))$ to be constant for their whole planning horizon (e.g., a week).
Thereby, the agents minimize the discomfort perceived today with a sensitivity $s^i(t)$ and the discomfort perceived in the remaining $T$ steps on the horizon with an average sensitivity $\bs$.
Finally, we assume agents to be conservative in terms of Karma: Given an initial Karma level $k^i(t)$, they will constrain their choices so that their Karma at the end of the horizon will not fall below a non-negative reference value $\kref^i$.
We explicitly account for path constraints (positive Karma at every time step) on the agent's Karma level only for the current decision, whilst assuming that in the remaining $T$ time-steps captured by the average behavior $\bar{y}^i$ they will be satisfied.
Formally, we get the following individual agent's problem:
\begin{problem}[Individual Agent's Problem]\label{prb:singleagent}
At time $t$, given the flows $x$ and prices $p$, respectively, a traveling agent with Karma level $k\geq 0$ and reference $\kref$, and sensitivity $s$ will choose her route as $y^\star$ resulting from
\par\nobreak\vspace{-10pt}
\begingroup
\allowdisplaybreaks
\begin{small}
\begin{subequations}\label{eq:singleAgentAverage}
	\begin{align}
	(y^\star,\by^\star) \in \argmin_{y\in\cY,\; \by\in\cbY} \;&s\cdot d(x)^\top y + T\cdot \bs\cdot d(x)^\top \by\\
	\text{s.t. } &k-p^\top y - T\cdot p^\top \by \geq k_\mathrm{ref}\\
	&p^\top y \leq k,
	\end{align}
\end{subequations}
\end{small}%
\endgroup
with $\cY = \{y\in\{0,1\}^n:\one^\top y = 1\}$ and $\cbY=\{\by\in[0,1]^n:\one^\top \by = 1\}$.
We define the set containing all points $y^\star$ solving~\eqref{eq:singleAgentAverage} as $\cY^\star(x,s,k,k_\mathrm{ref})\subseteq \cY$.
Non-traveling agents have $y^\star = \zero$.
\end{problem}
\noindent Note that a discount factor can be readily included by simply scaling the average sensitivity $\bs$ in the objective.

\subsection{Infinitely Many Agents Setting and  Wardrop Equilibrium}
In this paper, we consider the limit case whereby agents form a continuum with $M\to\infty$.
To describe the population, we use a notation similar to~\cite{GorokhBanerjeeEtAl2019}.
Thereby, we describe the distribution of the Karma level $k$ and reference $\kref$ in the population with the density function $\eta:\sR_+\times\sR_+\to\sR_+$.
This way, we can describe an infinite-agents population with $\rho(s)$ and $\eta(k,\kref)$.
For the infinite-agents setting, the Nash and Wardrop Equilibrium (WE) are identical~\cite{PaccagnanGentileEtAl2018} and can be defined as follows:
\begin{definition}[Wardrop Equilibrium]\label{def:WE}
$x^\mathrm{WE}\in[0,1]^n$ satisfying $\one^\top x^\mathrm{WE} = \Pgo$ is a WE if and only if there exist $y^\star(x^\mathrm{WE},s,k,k_\mathrm{ref})\in\cY^\star(x^\mathrm{WE},s,k,k_\mathrm{ref})$ so that
{\small
\begin{equation*}
    \int^{s_\mathrm{max}}_{s_\mathrm{min}} \int_{0}^\infty \int_{0}^\infty y^{\star}(x^\mathrm{WE},s,k,k_\mathrm{ref})\cdot \rho(s)\cdot\eta(k,\kref)\, \mathrm{d}s\,\mathrm{d}k\,\mathrm{d}\kref = x^\mathrm{WE}.
\end{equation*}}
\end{definition}
Informally, $x^\mathrm{WE}$ is a WE if the aggregate best-response based on assuming $x=x^\mathrm{WE}$ reconstructs the same vector $x^\mathrm{WE}$.
Observe that the agents' decision process can be interpreted as a model predictive control algorithm that implements solely the optimal decision for today $y^\star$, whilst discarding the optimal future decisions $\by^\star$.

Taking a static planning perspective, at each time-step $t$, we model the aggregate choices of the population with the WE $x^\mathrm{WE}(t)$.
The central operator's problem then is to choose the prices so that the daily WE $x^\mathrm{WE}(t)$ will converge to the system optimum $x^\star$.
\begin{problem}[Pricing Problem]\label{prb:prices}
Given a desired system optimum $x^\star$, the pricing problem consists of finding $p\in\sR^n$ so that $\lim_{t\to\infty} x^\mathrm{WE}(t) = x^\star$.
\end{problem}

\subsection{Discussion}
A few remarks are in order.
First, we use a static setting whereby on each day we compute the WE to model the users' behavior. This is in line with the mesoscopic perspective of our study.
Second, we assume the agents to be rational and to share the discomfort function $d(x)$ and sensitivity probability density function $\rho(s)$, but differentiate them with regard to their daily level of sensitivity $s^i(t)$ and their reference Karma-level $k_\mathrm{ref}^i$.

\section{Two-arcs Pricing Problem}\label{sec:twoarcs}
For the remainder of this paper, we will focus on the parallel network with two arcs shown in Fig.~\ref{fig:ODwithMultiarcs}.
This model can represent a daily commute routine where users can choose between two options, a fast (or comfortable) route, and one that is slower (or less comfortable). We leave the study of parallel networks with more arcs and more general transportation networks to future research endeavors.
We assume, without loss of generality, that at the desired socially-optimal solution $x^\star$ it holds that $d_1(x_1^\star)<d_2(x_2^\star)$.\footnote{If it were to hold $d_2(x_2^\star)<d_1(x_1^\star)$, given the monotonocity properties of $d(x)$, we could simply swap the arcs' labeling and recover this problem.}
Therefore, we introduce a price $p_1>0$ for arc~1 and a reward $r_2 = -p_2>0$ for arc 2.
Crucially, given a desired solution $x^\star$, a necessary condition required to solve Problem~\ref{prb:prices}, is that the total Karma must be conserved in the steady state. 
Hence, the steady-state prices must satisfy $p^\top x^\star = 0$, implying that
\par\nobreak\vspace{-5pt}
\begingroup
\allowdisplaybreaks
\begin{small}
\begin{equation}\label{eq:prices}
p_1 = r_2\cdot \frac{x_2^\star}{x_1^\star}.
\end{equation}
\end{small}%
\endgroup
Since scaling the prices' $(p_1,p_2)$ with a common factor would not change the set of equilibria, \eqref{eq:prices} automatically delivers the only possible prices that could solve Problem~\ref{prb:prices}.
Therefore, in the two-arcs setting under consideration, there is no control freedom when choosing the steady-state prices.
However, we will show in the remainder of this paper that choosing the prices as in~\eqref{eq:prices} will make the aggregate behavior of the agents described by Problem~\ref{prb:singleagent} converge to the desired system optimum resulting from Problem~\ref{prb:SO}.

\section{Best Response Strategy}\label{sec:bestresponse}
This section derives the best response strategy for an agent with Karma-level $k$, reference level $\kref$ and daily sensitivity $s$.
We begin with a technical assumption ensuring feasibility of the problem at hand.
\begin{assump}[Feasible Desired Flow]\label{ass:meaningfulFlows}
	The desired mesoscopic flow satisfies $x^\star>0$ and $x_1^\star/x_2^\star\in[1/T,T]$, so that $r_2/p_1\in[1/T,T]$.
\end{assump}
\noindent This assumption is needed because, if the ratio between price $p_1$ and reward $r_2$ was too small or too large, there would not always exist a $\by\in\cbY$ so that $p^\top(y+T\cdot \by)$ is zero for all $y\in\cY$ and, therefore, a \emph{Karma-sustained} planning would be infeasible.
However, it is reasonable to consider a long-enough horizon $T$, so that the latter requirement would not be limiting.
\begin{figure}[t]
	\includegraphics[width=\columnwidth]{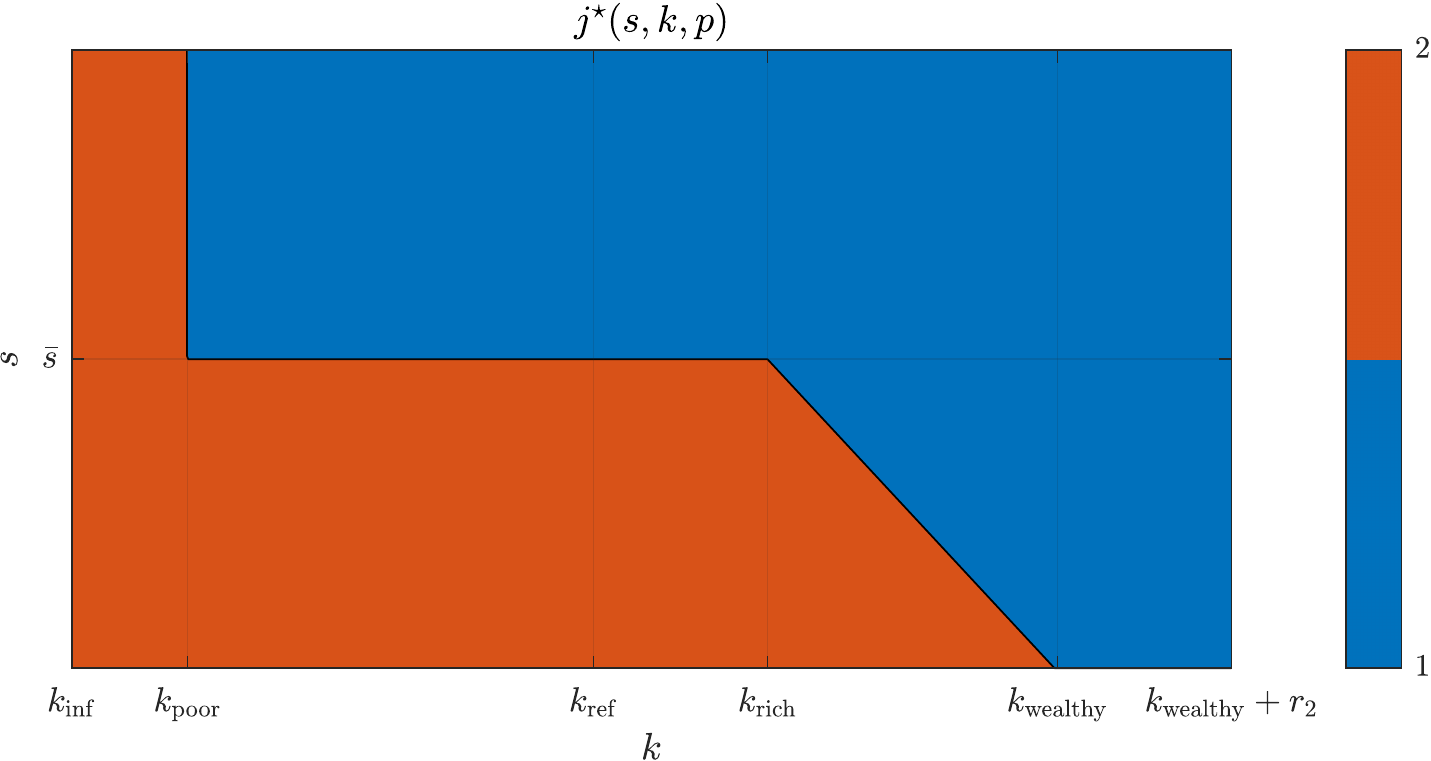}
	\caption{The best response strategy~\eqref{eq:bestResponse} resulting from the solution of Problem~\ref{prb:singleagent} for $d_1(x)<d_2(x)$.}
	\label{fig:bestResponse}
\end{figure}
    
The solution of Problem~\ref{prb:singleagent}, namely the agents' best response strategy, can be computed in closed-form as shown in the following theorem.
\begin{theorem}[Best Response Strategy]\label{thm:brs}
	We define $k_\mathrm{inf}=\max\{0,k_\mathrm{ref} - (T+1)\cdot r_2\}$, $k_\mathrm{poor} = \max\{p_1,k_\mathrm{ref} + p_1 - T\cdot r_2\}$,  \mbox{$k_\mathrm{rich} = k_\mathrm{ref} +T\cdot p_1 - r_2$}, and $k_\mathrm{wealthy} = k_\mathrm{ref} +(T+1)\cdot p_1$. Consider a player with Karma $k\geq k_\mathrm{inf}$, sensitivity $s$, Karma reference $k_\mathrm{ref}$, and given prices $p$.  Given $x$ so that for the discomfort it holds $d_1(x_1)<d_2(x_2)$, the best response strategy of a traveling agent in terms of which arc $j$ to choose, so that $y^\star_{j^\star}=1$ and 0 otherwise, is
	\par\nobreak\vspace{-5pt}
    \begingroup
    \allowdisplaybreaks
    \begin{small}
	\begin{equation}\label{eq:bestResponse}
	j^\star = \begin{cases}
	2 & \text{if } k \in[k_\mathrm{inf},k_\mathrm{poor})\\
	\begin{cases}
	1 & \text{if } s> \bs\\
	2 & \text{if } s< \bs
	\end{cases} & \text{if } k \in[k_\mathrm{poor},k_\mathrm{rich})\\
	\begin{cases}
	1 & \text{if } s> \bs\cdot\frac{\kwealthy-k}{p_1+r_2}\\
	2 & \text{if } s< \bs\cdot\frac{\kwealthy-k}{p_1+r_2}
	\end{cases} & \text{if } k \in[k_\mathrm{rich},k_\mathrm{wealthy})\\
	1 & \text{if } k \geq k_\mathrm{wealthy}.
	\end{cases}
	\end{equation}
	\end{small}
	\endgroup
	Moreover, given $x$ so that $d_1(x_1)=d_2(x_2)$, the best response strategy of a traveling agent is
	\par\nobreak\vspace{-5pt}
    \begingroup
    \allowdisplaybreaks
    \begin{small}
	\begin{equation}\label{eq:bestResponseEq}
	j^\star \begin{cases}
	=2 & \text{if } k \in[k_\mathrm{inf},k_\mathrm{poor})\\
	\in\{1,2\} & \text{if } k \geq k_\mathrm{poor}.
	\end{cases}
	\end{equation}
	\end{small}%
	\endgroup
	Finally, given $x$ so that $d_1(x_1)>d_2(x_2)$, the best response strategy of a traveling agent is
	\par\nobreak\vspace{-5pt}
    \begingroup
    \allowdisplaybreaks
    \begin{small}
	\begin{equation}\label{eq:bestResponsed1worse}
	j^\star =2 \quad \forall k \geq k_\mathrm{inf}.
	\end{equation}
	\end{small}%
	\endgroup
\end{theorem}

\noindent The proof can be found in
\ifextendedversion
Appendix~\ref{app:proof}.
\else
the extended version of this paper~\cite{SalazarPaccagnanEtAl2021}.
\fi
Fig.~\ref{fig:bestResponse} depicts the best response strategy~\eqref{eq:bestResponse}.
As shown in the proof, Problem~\ref{prb:singleagent} is infeasible for $k< k_\mathrm{inf}$. Therefore, we will always consider the case whereby the Karma of any agent is initialized above $k_\mathrm{inf}$.
Moreover, under policy~\eqref{eq:bestResponse} the set $[k_\mathrm{inf},k_\mathrm{wealthy}+r_2)$ is positively invariant and attractive from above, so that it will be $k\ge k_\mathrm{inf}$ for all future times.
Finally, we observe that the best response strategies~\eqref{eq:bestResponse}--\eqref{eq:bestResponsed1worse} are independent of the quantitative discomfort level; they depend only on the prices, the current level of Karma $k$, and the sensitivity $s$.

\subsection{Wardrop Equilibrium}\label{subsec:brsdisc}
One can verify that, when prices are chosen according to~\eqref{eq:prices}, there exists no WE satisfying $d_1(x_1)>d_2(x_2)$. This is due to the fact that, when $d_1(x_1)>d_2(x_2$), the agent's best response would entail choosing arc~2, thus leading to $d_1(0)<d_2(\Pgo)$.
Finally, we define $\bx$ so that $d_1(\bx_1) = d_2(\bx_2)$, whereby, given the monotonicity properties of $d(x)$, we have that $d_1(x_1) < d_2(x_2)$ for all $x_1<\bx_1$.
In the following lemma, we show that a WE always exists, and is characterized by $d_1(x_1^\mathrm{WE}) \leq d_2(x_2^\mathrm{WE})$.
\begin{lemma}[Wardrop Equilibrium for Two Arcs]\label{lem:WE2}
Let $\tilde{x}_1<\bx_1$, i.e., $d_1(\tilde{x}_1)<d_2(\tilde{x}_2)$, and define the mesoscopic flow resulting from the best response~\eqref{eq:bestResponse} $y^\star(\tilde{x},s,k,k_\mathrm{ref})\in\cY^\star(\tilde{x},s,k,k_\mathrm{ref})$ as
\par\nobreak\vspace{-5pt}
    \begingroup
    \allowdisplaybreaks
    \begin{small}
\begin{equation}\label{eq:xbr}
    x^\mathrm{BR}=\int^{s_\mathrm{max}}_{s_\mathrm{min}} \int_{0}^\infty \int_{0}^\infty y^{\star}(\tilde{x},s,k,k_\mathrm{ref})\cdot \rho(s)\cdot\eta(k,\kref)\, \mathrm{d}s\,\mathrm{d}k\,\mathrm{d}\kref.
\end{equation}
\end{small}%
\endgroup
A WE exists and is characterized by $x^\mathrm{WE} = x^\mathrm{BR}$ if $x^\mathrm{BR}_1 < \bx_1$, and/or $x^\mathrm{WE} = \bx$.
\end{lemma}
\noindent The proof can be found in
\ifextendedversion
Appendix~\ref{app:proof3}.
\else
the extended version of this paper~\cite{SalazarPaccagnanEtAl2021}.
\fi
We define the WE at $\bx$ as \emph{uncontrolled}, since it corresponds to the WE resulting when no pricing is applied. In fact, it exists when the population has too much Karma and can therefore easily afford the first arc.

For the remainder of this paper we will assume that for the discomfort functions it holds $d_1(x_1)<d_2(x_2)$ for all admissible $x$.
This way, in addition to Lemma~\ref{lem:WE2}, we can prove the existence of a unique WE with $d_1(x_1^\mathrm{WE}) < d_2(x_2^\mathrm{WE})$ for any Karma distribution $\eta(k,\kref)$, directly stemming from the best response strategy~\eqref{eq:bestResponse}.
This assumption enables the formal analysis of the mesoscopic aggregate behavior presented in Section~\ref{sec:mesoscopic} below.
However, we believe this assumption can be lifted, since by the prices definition~\eqref{eq:prices}---i.e., $p_1x_1^\star-r_2x_2^\star=0$---and the fact that $x_1^\star < \bx_1$, it holds that $p^\top\bx>0$. Therefore, the uncontrolled WE at $\bx$ is not sustainable in time, as the population would continuously lose Karma, until eventually reaching a distribution with WE \mbox{$d_1(x_1^\mathrm{WE}) < d_2(x_2^\mathrm{WE})$} converging to $x^\star$.
We corroborate our conjecture with numerical simulations in Section~\ref{sec:results}, whilst leaving a formal mathematical proof to an extended version of this paper.

\section{Mesoscopic Average Behavior}\label{sec:mesoscopic}
This section studies the dynamics of the aggregate behavior of the population.
In the case under consideration, the WE always results from the individual agent's best response strategy~\eqref{eq:bestResponse}, i.e., it corresponds to $x^\mathrm{BR}$ from~\eqref{eq:xbr} in Lemma~\ref{lem:WE2}.
Since the positively invariant set of~\eqref{eq:bestResponse} $k\in[k_\mathrm{inf},\kwealthy+r_2)$ is attractive, we assume each agent to start with a level of Karma already in it.
We first define the fraction of agents with a certain level of Karma:
\par\nobreak\vspace{-5pt}
    \begingroup
    \allowdisplaybreaks
    \begin{small}
\begin{equation}\label{eq:probKarma}
\begin{aligned}
P_\mathrm{poor} &= P(k\in[k_\mathrm{inf},\kpoor))\\
P_\mathrm{ok} &= P(k\in[\kpoor,\krich))\\
P_\mathrm{rich} &= P(k\in[\krich,\kwealthy))\\
P_\mathrm{wealthy} &= P(k\in[\kwealthy,\kwealthy+r_2))\\
\end{aligned}
\end{equation}
\end{small}%
\endgroup
Next, given the probability density function of the sensitivity $\rho(s)$ and $\hat{s}\in[s_\mathrm{min},s_\mathrm{max}]$, we define
\par\nobreak\vspace{-5pt}
    \begingroup
    \allowdisplaybreaks
    \begin{small}
\begin{equation}
\begin{aligned}
P_\mathrm{chill}(\hat{s}) & =P(s< \hat{s}) = \int_{s_\mathrm{min}}^{\hat{s}}\rho(s)\,\mathrm{d}s\\
P_\mathrm{rush}(\hat{s}) & = P(s> \hat{s}) = 1-P_\mathrm{chill}(\hat{s}) = \int_{\hat{s}}^{s_\mathrm{max}}\rho(s)\,\mathrm{d}s.
\end{aligned}
\end{equation}
\end{small}%
\endgroup
Overall, these six measures represent the fraction of agents present in one of the regions shown in Fig.~\ref{fig:bestResponse}, and can be used to compute the average population response.
For the sake of simplicity, but without loss of generality, from now on we assume \mbox{$r_2=-p_2\geq p_1>0$}, and $k_\mathrm{ref} +p_1-T\cdot r_2\geq p_1$, i.e., \mbox{$k_\mathrm{poor} = k_\mathrm{ref} + p_1-T\cdot r_2$}. The cases with $k_\mathrm{poor} = p_1$ and/or $r_2\leq p_1$ can be studied with identical arguments and lead to the same result.
Furthermore, we assume $p_1$ and $r_2$ to be commensurate, and scale them to be integer and co-prime (in practice, it is possible to round the prices after a suitable scaling without appreciably affecting the results, as shown in Section~\ref{sec:results} below). We define \mbox{$i= k-k_\mathrm{ref} + T\cdot r_2+1$}, quantize the probability distributions~\eqref{eq:probKarma} as
\par
\begin{figure}[t]
\includegraphics[width=\columnwidth]{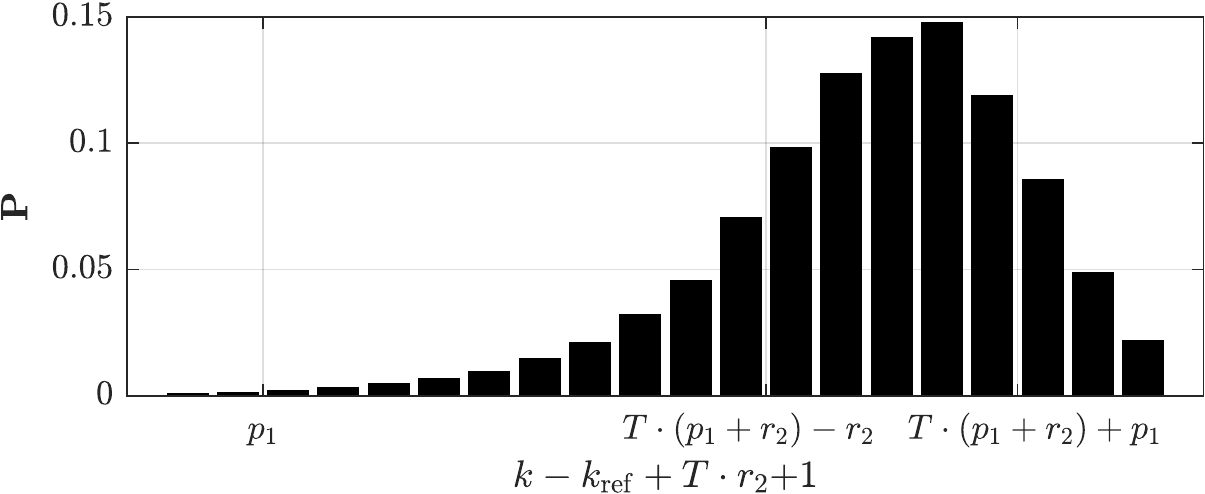}
\caption{The steady-state distribution in relative Karma levels for an exponentially distributed $s$.}
\label{fig:PDistribution}
\end{figure}

\nobreak\vspace{-5pt}
    \begingroup
    \allowdisplaybreaks
    \begin{small}
\begin{equation}
\begin{aligned}
&P_\mathrm{poor} = \summe{i=1}{p_1} \vP_{\mathrm{poor}}(i),
&&P_\mathrm{ok} = \summe{i=1}{(T-1)\cdot(p_1+r_2)} \vP_{\mathrm{ok}}(i),\\
&P_\mathrm{rich} = \summe{i=1}{p_1+r_2} \vP_{\mathrm{rich}}(i),
&&P_\mathrm{wealthy} = \summe{i=1}{r_2} \vP_{\mathrm{wealthy}}(i),\\
\end{aligned}
\end{equation}
\end{small}%
\endgroup
and stack them in the vector $\vP=(\vP_\mathrm{poor}^\top,\vP_\mathrm{ok}^\top,\vP_\mathrm{rich}^\top,\vP_\mathrm{wealthy}^\top)^\top$,
as shown in Fig.~\ref{fig:PDistribution}.
\begin{figure*}[t]
	\begin{equation}\label{eq:evolution}
	\footnotesize
		\begin{aligned}
			&\vP(i)^+ = \Pho\cdot \vP(i) + \Pgo\cdot \Pru(\bs)\cdot \vP(i+p_1) &&\forall i \in [1,r_2]\\
			&\vP(i)^+ = \Pho\cdot \vP(i) + \Pgo\cdot (\vP(i-r_2)+\Pru(\bs)\cdot \vP(i+p_1)) && \forall i \in[r_2+1,r_2+p_1]\\
			&\vP(i)^+ = \Pho\cdot \vP(i) + \Pgo\cdot(\Pch(\bs)\cdot \vP(i-r_2)+\Pru(\bs)\cdot \vP(i+p_1)) && \forall i \in[r_2+p_1+1,(T-1)\cdot(p_1+r_2)]\\
			&\vP(i)^+ = \Pho\cdot \vP(i) + \Pgo\cdot(\Pch(\bs)\cdot \vP(i-r_2)+ \Pru\left(\bs\cdot\frac{T\cdot(p_1+r_2) +1-i}{p_1+r_2}\right)\vP(i+p_1)) && \forall i \in[(T-1)\cdot(p_1+r_2)+1,T\cdot(p_1+r_2)]\\
			&\vP(i)^+ = \Pho\cdot \vP(i) + \Pgo\cdot(\Pch\left(\bs\cdot\frac{(T+1)\cdot(p_1+r_2) +1-i}{p_1+r_2}\right)\cdot \vP(i-r_2)+ \vP(i+p_1)) && \forall i \in[T\cdot(p_1+r_2)+1,T\cdot p_1+(T+1)\cdot r_2]\\
			&\vP(i)^+ = \Pho\cdot \vP(i) + \Pgo\cdot\Pch\left(\bs\cdot\frac{(T+1)\cdot(p_1+r_2) +1-i}{p_1+r_2}\right)\cdot \vP(i-r_2) && \forall i \in[T\cdot p_1+(T+1)\cdot r_2+1,(T+1)\cdot (p_1+r_2)]\\
		\end{aligned}
	\end{equation}
\begin{subequations}
\footnotesize
\begin{align}
&A_\mathrm{chill}(i,j) = \begin{cases}
1 & \text{if } i\in [r_2+1,r_2+p_1]\wedge j = i-r_2\\
P_\mathrm{chill}(\bs) & \text{if } i\in [p_1+r_2+1,T\cdot(p_1+r_2)]\wedge j = i-r_2\\
P_\mathrm{chill}\left(\bs\cdot\frac{(T+1)\cdot(p_1+r_2) +1-i}{p_1+r_2}\right) & \text{if } i\in [T\cdot(p_1+r_2)+1,(T+1)\cdot(p_1+r_2)]\wedge j = i-r_2\\
0 & \text{otherwise}
\end{cases}\label{subeq:achill}\\
&A_\mathrm{rush}(i,j) = \begin{cases}
P_\mathrm{rush}(\bs)& \text{if } i\in [1,(T-1)\cdot(p_1+r_2)]\wedge j = i+p_1\\
P_\mathrm{rush}\left(\bs\cdot\frac{T\cdot(p_1+r_2) +1-i}{p_1+r_2}\right)& \text{if } i\in [(T-1)\cdot(p_1+r_2)+1,T\cdot(p_1+r_2)]\wedge j = i+p_1\\
1 & \text{if } i\in [T\cdot(p_1+r_2)+1,T\cdot p_1+(T+1)\cdot r_2]\wedge j =  i+p_1\\
0 & \text{otherwise}
\end{cases}\label{subeq:arush}
\end{align}
\end{subequations}
\hrulefill
\end{figure*}
The probability vector $\vP$ is the quantized version of $\eta(k,\kref)$, and evolves in line with the best response strategy~\eqref{eq:bestResponse} as detailed in~\eqref{eq:evolution} on page~\pageref{eq:evolution}.
In matrix form, it can be written as
\par
\begin{figure}[t]
	\centering
	\includegraphics[width=0.9\columnwidth]{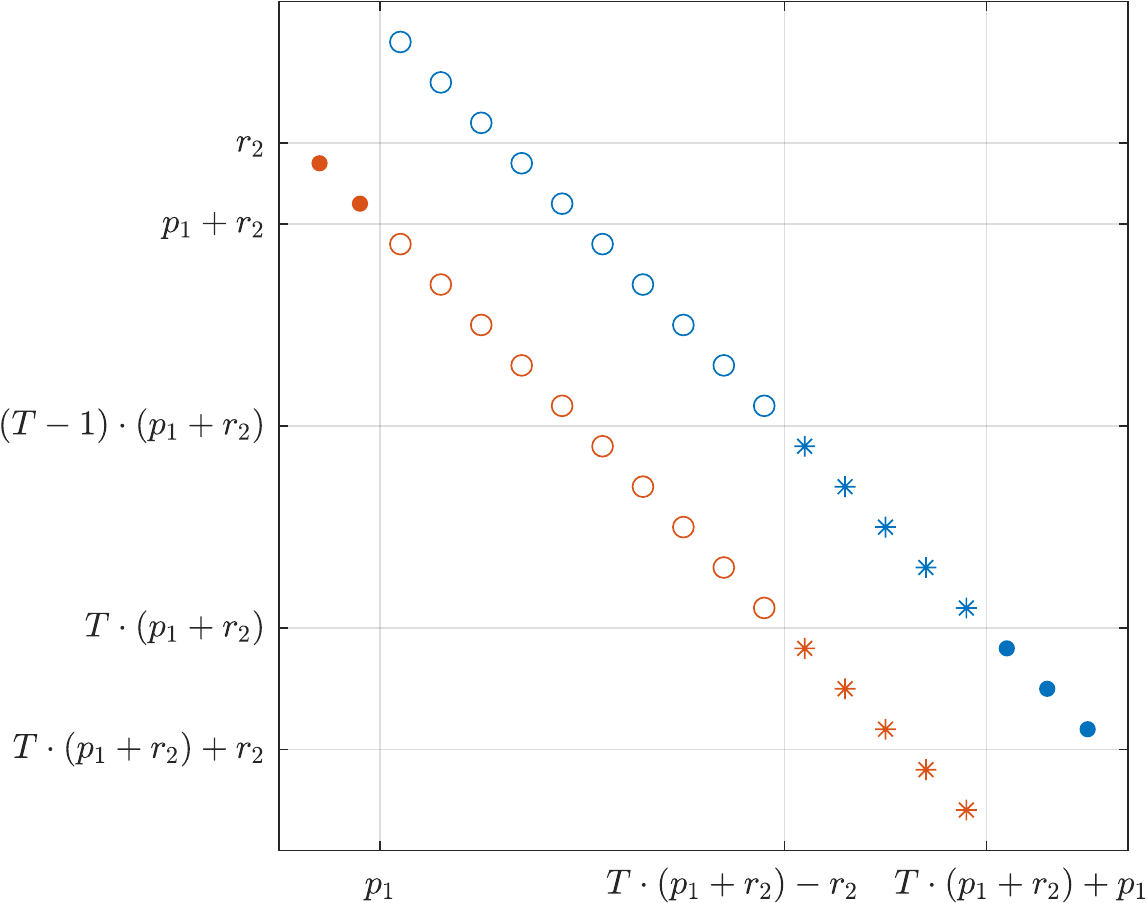}
	\caption{Sparsity pattern of the matrix $A_\mathrm{go}$. The probability of choosing the first arc paying $p_1$ is denoted in blue, whilst the probability of choosing the second arc receiving $r_2$ in red. In this example, we chose $p_1=2$, $r_2=3$ and $T=3$. Full dots represent 1-entries, whilst empty dots denote $\Pru(\bs)$ (blue) and $\Pch(\bs)$ (red). Stars represent $\Pru(.)$ (blue) and $\Pch(.)$ (red) evaluated as in the third and second condition of~\eqref{subeq:achill} and~\eqref{subeq:arush}, respectively. Since $\Pru+\Pch=1$, column-stochasticity of $A_\mathrm{go}$ can be determined by inspection.}
	\label{fig:MatrixA}
\end{figure}

\nobreak\vspace{-5pt}
    \begingroup
    \allowdisplaybreaks
    \begin{small}
\begin{equation}\label{eq:dlti}
\vP^+=A \vP,
\end{equation}
\end{small}%
\endgroup
where $A= \Pho\cdot I + \Pgo\cdot A_\mathrm{go}$ is a non-negative, square and column-stochastic matrix, i.e., $A^\top 
\one = \one$.
Furthermore, $A$ is primitive if $\Pho>0$.
Since $A$ is non-negative, for any $\vP\geq \zero$, it holds that $\vP^+\geq \zero$. Moreover, being $A$ column-stochastic, we have that $\one^\top \vP^+= \one^\top A \vP= \one^\top \vP$~\cite{Bullo2018}.
The sparsity pattern of $A_\mathrm{go}=A_\mathrm{chill}+A_\mathrm{rush}$ is shown in Fig.~\ref{fig:MatrixA}, whereby the red dots represent the probability for a traveling agent with Karma deviation \mbox{$k = i +k_\mathrm{ref}+ T\cdot r_2-1$} to choose the slow arc and are defined as in~\eqref{subeq:achill}, whilst the blue dots represent the probability to choose the fast arc and are defined as in~\eqref{subeq:arush}.
The system has discrete linear time-invariant dynamics.
Its $\omega$-limit sets are either fixed points or limit cycles. Such $\omega$-limit sets can be found by eigenvalue analysis as we do below, showing that limit cycles can only appear if \mbox{$\Pho = 0$}.
Since $\vP\geq \zero$ is a probability distribution, it must hold $\one^\top \vP=1$. Hence, the trivial equilibrium $\vP=\zero$ is not admissible. As the equilibrium is defined as $\Pe= A \Pe$, it must be spanned by the eigenvector related to the 1-eigenvalue of $A$.

With this in mind, we can now show that in the large population limit, the equilibrium in Karma distribution $\Pe$ is globally asymptotically stable and corresponds to the desired system optimum~$x^\star$.
\begin{theorem}[Globally Asymptotically Stable and Optimal Equilibrium]\label{thm:eq}
    Given the prices~\eqref{eq:prices} and a population of agents acting in line with~\eqref{eq:bestResponse} with $\Pho>0$, the equilibrium of the Karma distribution dynamics~\eqref{eq:dlti} $\Pe\geq 0$ with \mbox{$\one^\top\Pe=1$} is globally asymptotically stable and its resulting flows correspond to the system optimum~$x^\star$.
\end{theorem}
\noindent The proof can be found in
\ifextendedversion
Appendix~\ref{app:proof2}.
\else
the extended version of this paper~\cite{SalazarPaccagnanEtAl2021}.
\fi
This theorem shows that the prices resulting from Karma conservation arguments indeed solve Problem~\ref{prb:prices}.
Note that equivalent results can be obtained by leveraging the geometric ergodicity properties of the Markov chain $\vP$ through straightforward application of Doeblin's theorem~\cite{Doeblin1937}.
\section{Numerical Results}\label{sec:results}
This section presents the results obtained via numerical simulations for the following case-studies: First, a scenario where the social cost corresponds to the sum of the agents' cost, i.e., $d(x)^\top x$ as in standard routing settings~\cite{BrownMarden2017}.
There we consider the case whereby on average $\Pho=5\%$ of the population does not travel and the limit case where all agents travel every day, i.e., $\Pho = 0$.
Second, we study the case where the societal cost does not correspond to the sum of the users' costs. For all the scenarios we consider $M=1000$ agents with a horizon of a week with \mbox{$T=6$}, and sample their daily sensitivity from an exponential distribution defined on $\sR_+$.
In line with our theoretical findings, the case where the discomfort functions satisfy $d_1(x_1)<d_2(x_2)$ for any admissible $x$ resulted in the expected convergence of the aggregate behavior to the system optimum for any of the scenarios mentioned.
For the sake of brevity, we omit such results and focus on the more interesting case whereby $d_1(x_1)>d_2(x_2)$ for $x_1>\bx_1$, and the uncontrolled WE at $\bar{x}$ exists.
Specifically, we model the discomfort as a travel-time Bureau of Public Roads (BPR) function~\cite{BPR1964}
\par\nobreak\vspace{-5pt}
    \begingroup
    \allowdisplaybreaks
    \begin{small}
\begin{equation}
    d(x_j) = d_{0,j}\cdot\left(1+\alpha\cdot\left(x_j/\kappa_j\right)^\beta\right),
\end{equation}
\end{small}%
\endgroup
with $d_0=(1,2)^\top$, $\kappa = (1/2,2/3)^\top$, $\alpha = 0.15$ and $\beta=4$, for which $\bx_1 = 0.80$ for both values of $\Pho$.
We simulate each day by iteratively computing the Nash equilibrium (approximating the infinite-agents WE) resulting when each agent is solving Problem~\ref{prb:singleagent} (which can be efficiently solved as a linear program).
The computation of the Nash equilibrium always needed only a couple of iterations.

\subsection{The Social Cost is the Sum of the Agents' Cost}
\begin{figure}[t]
	\includegraphics[width=\columnwidth]{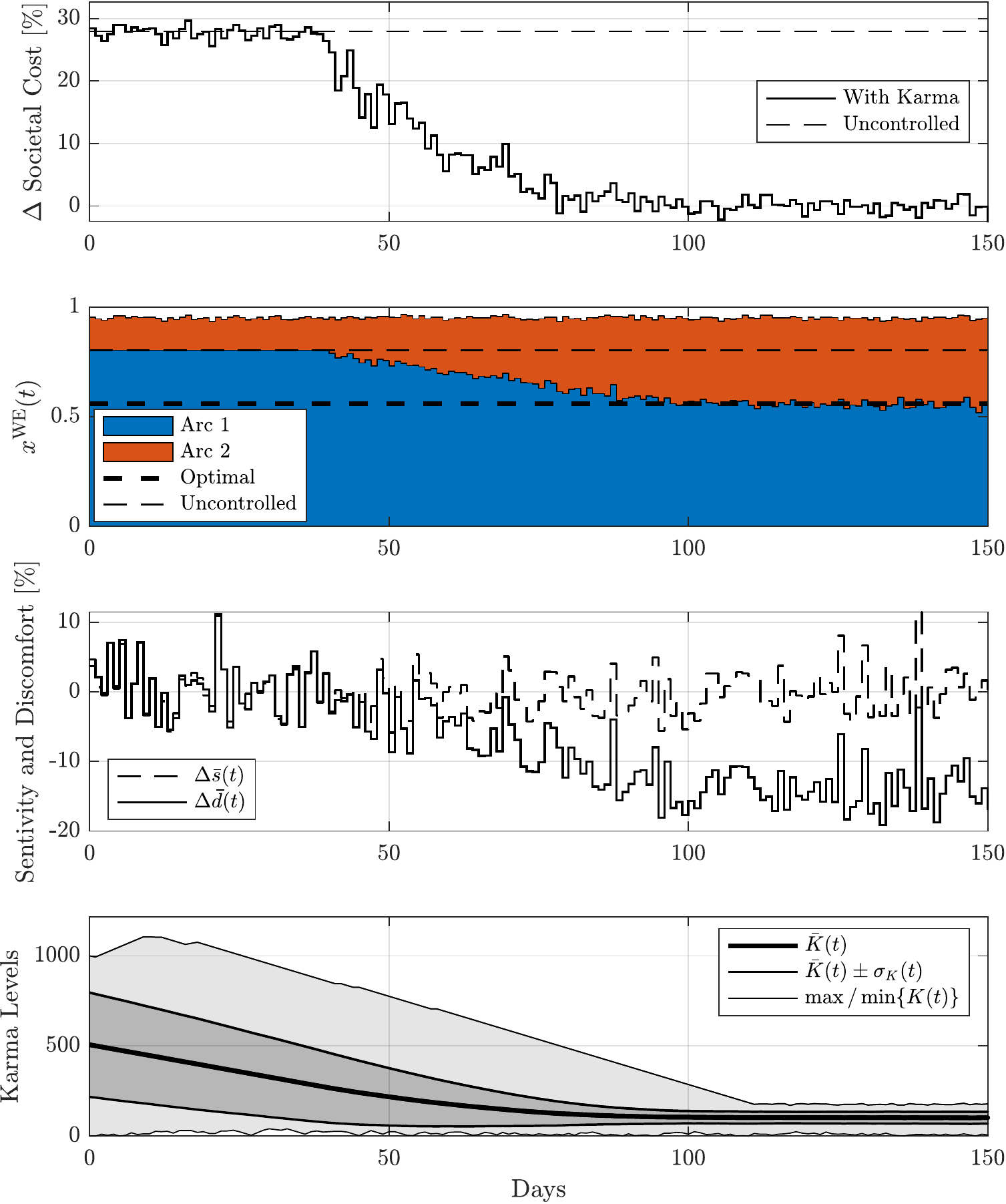}
	\caption{
		Societal cost, mesoscopic flows, relative average sensitivity and discomfort deviation, and Karma distribution for a population with \mbox{$\Pho = 5\%$} and very large initial Karma levels.
	}
	\label{fig:sim}
\end{figure}
Setting $\Pho=5\%$, the system optimum is $x^\star=(0.56,0.39)^\top$, for which we design prices according to~\eqref{eq:prices} as $p=(10,-14)^\top$, rounding them as mentioned in Section~\ref{sec:mesoscopic} above.
We initialize the Karma reference levels $\kref^i$ from a uniform distribution defined between 0 and~100, whilst initializing the Karma levels $k^i(0)$ between 0 and 500, so that an extremely large fraction of agents starts with $k$ above $\kwealthy$ and thus $\bx$ is the only possible WE.
As expected, Fig.~\ref{fig:sim} shows that providing the agents with too much Karma will indeed result in the uncontrolled WE.
However, as the average Karma level of the population $\bK(t)$ shown in the fourth subplot is depleted, the system-level behavior and cost converge very close to the system optimum with an average relative societal cost difference below 0.1\%.
We measure the perceived discomfort of the single agents $s^i(t)\cdot d(x(t))^\top y^i(t)$ and average it over the population. The third subplot of Fig.~\ref{fig:sim} compares it with the average discomfort that would be perceived by the agents if they were allocated in a random and sensitivity-unaware fashion to the same flows as
\par\nobreak\vspace{-5pt}
\begingroup
\allowdisplaybreaks
\begin{small}
\begin{equation}\label{eq:deltad}
    \Delta \bar{d}(t) = \frac{\sum_{i=1}^{1000} s^i(t)\cdot d(x(t))^\top y^i(t) - \bs\cdot d(x(t))^\top y^i(t)}{\sum_{i=1}^{1000}\bs\cdot d(x(t))^\top y^i(t)},
\end{equation}
\end{small}%
\endgroup
which exhibits a behavior that is similar to the relative deviation of the daily average sensitivity from the distribution's mean \mbox{$\Delta \bar{s}(t) = \sum_{i=1}^{1000}\frac{s^i(t)-\bs}{1000\cdot \bs}$}.
In line with the goal of our framework, while converging to the system optimum, the agents perceive a discomfort about 14\% lower compared to the case where the agents would be randomly allocated in a system-optimal but urgency-unaware fashion.

\begin{figure}[t]
	\includegraphics[width=\columnwidth]{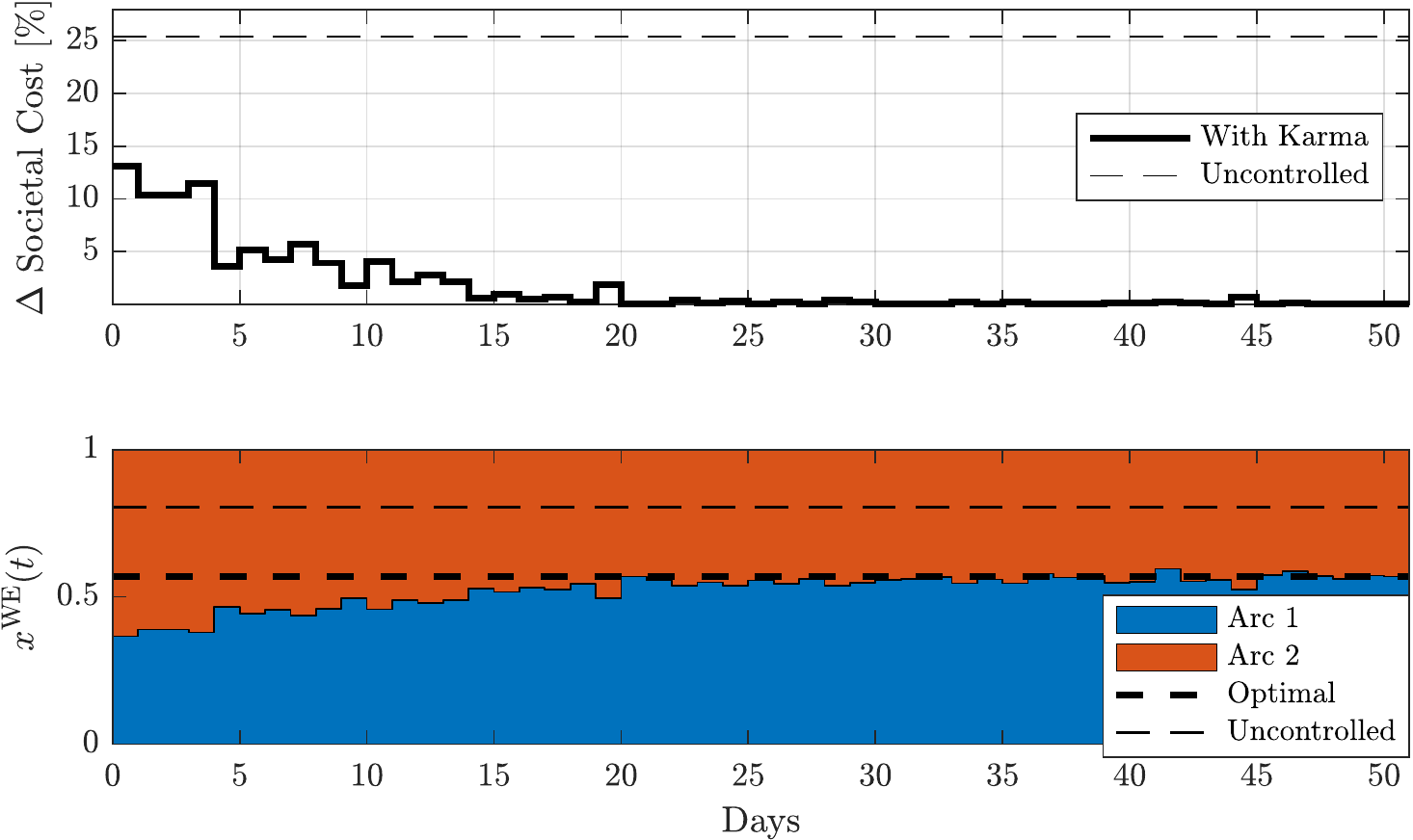}
	\caption{
		Societal cost and mesoscopic flows for $\Pho = 0$.
	}
	\label{fig:simP0}
\end{figure}

Next, we study the limit case where all agents are traveling, i.e., $\Pho=0$. Fig.~\ref{fig:simP0} shows the results obtained in this scenario with system-optimal solution $x^\star=(0.57,0.43)^\top$ for which we design again the prices via~\eqref{eq:prices} and round them to $p=(10,-13)^\top$.
In this case, we initialize the Karma initial and reference values
between 0 and 100.
Interestingly, the proposed scheme seems to work well also for this limit case that could admit periodic solutions---as the matrix $A$ would no longer be primitive but only row-stochastic.
Again, the population behavior converges very close to the system-optimum with a relative deviation of about 0.2\% and an average $\Delta \bd$ of about -14\%. Since in this scenario the number of users traveling every day is constant, the optimal societal cost is never outperformed.
Overall, these results prompt us to study in more detail also the convergence properties of this limit case with $\Pho = 0$.

\subsection{The Social Cost is not the Sum of the Agents' Cost}
\begin{figure}[t]
	\includegraphics[width=\columnwidth]{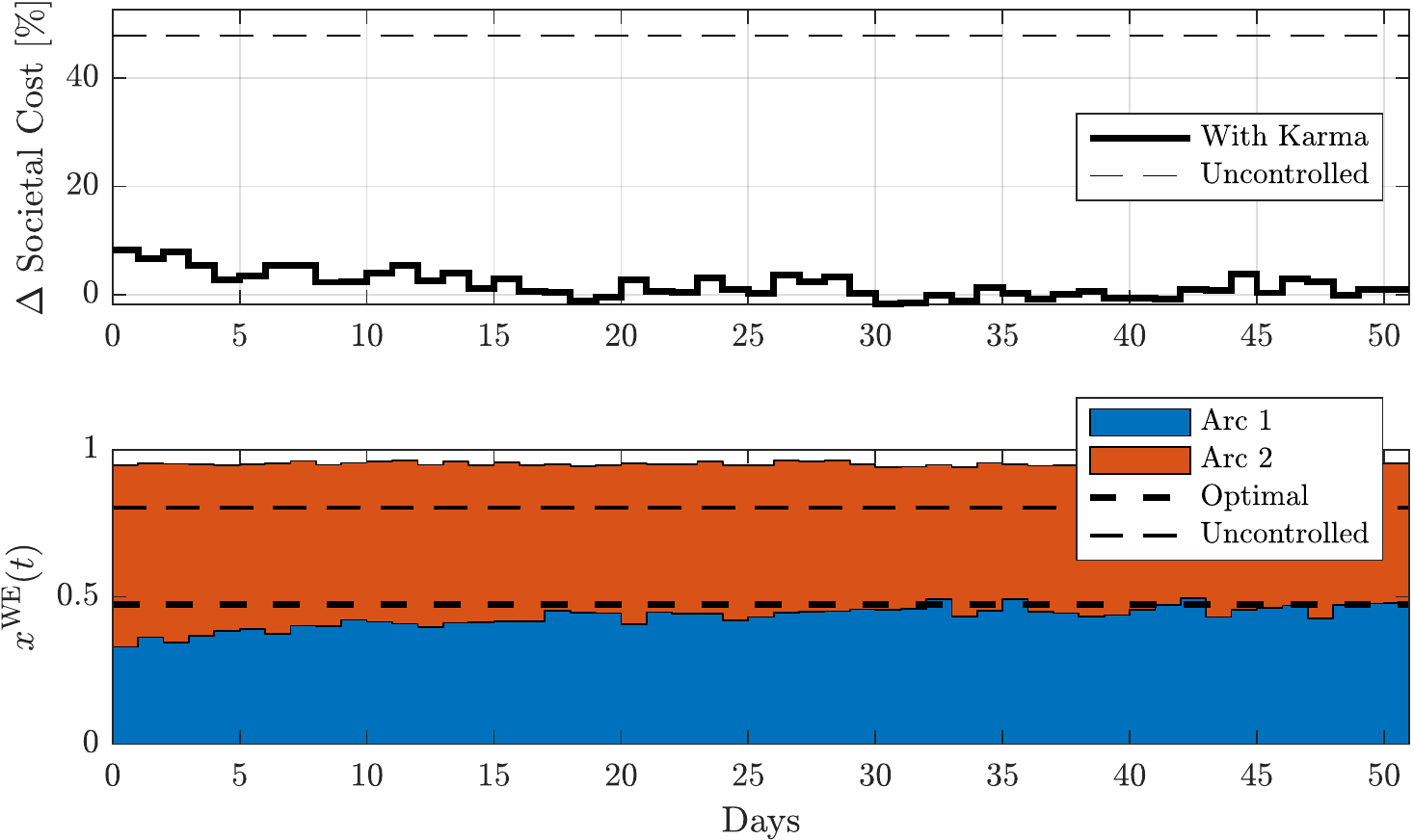}
	\caption{
		Societal cost and mesoscopic flows for $c(x) = x$ and $\Pho=5\%$.
	}
	\label{fig:simCnD}
\end{figure}
The proposed framework enables to steer the behavior of the population to any average choice satisfying Assumption~\ref{ass:meaningfulFlows}. In this case, we set the societal cost as $c(x)=x$, leading to the symmetric optimal flows $x^\star=(0.5,0.5)^\top$ for which the prices resulting from~\eqref{eq:prices} can be chosen as $p=(10,-10)^\top$.
Fig.~\ref{fig:simCnD} shows that, despite different societal and users' objectives and a very high price of anarchy for the uncontrolled case, the mesoscopic flows converge to the system optimum.
Finally, also in this scenario, our framework enables agents to significantly improve the perceived discomfort by about 20\% w.r.t.\ a system-optimal but random allocation, whilst aligning their behavior with the desired system optimum.
\section{Conclusion}\label{sec:conclusion}
This paper explored the application of artificial currencies to route self-interested agents in a system-optimal fashion whilst accounting for their temporal needs.
Specifically, we instantiated a repeated game in a static setting whereby each day traveling agents choose whether to cross the most comfortable route for a price or to receive a reward for traveling a less comfortable route.
For a two-parallel-arcs network we showed that a static pricing choice resulting from simple Karma-conservation arguments guarantees the mesoscopic average behavior to converge to the system optimum, significantly reducing the societal costs w.r.t.\ the uncontrolled equilibrium.
What is more, our scheme is fairly simple and does not rely on any auction mechanism for allocation, but leaves full freedom of choice to the agents as long as their Karma level is non-negative. As a result, it enables a considerable reduction of the perceived discomfort with respect to a random but optimal allocation.
In practice, our numerical results were in line with our findings in all the scenarios studied: Our scheme reached a societal cost less than 1\% higher than the system optimum, while significantly reducing the average perceived discomfort by 14-20\%.

This work can be extended as follows.
We would like to study more realistic network structures, such as more general transportation graphs with different origin-destination pairs.
Furthermore, we would like to devise learning-based control algorithms to adapt the prices in real-time and implement them within high-fidelity simulation environments.

\ifextendedversion
\begin{appendices}
\section{Proof of Theorem~\ref{thm:brs}}\label{app:proof}
Before solving Problem~\ref{prb:singleagent}, we reformulate it as
\par\nobreak\vspace{-5pt}
    \begingroup
    \allowdisplaybreaks
    \begin{small}
	\begin{subequations}
		\begin{align}
		\min_{y,\by}\; & d^\top\left(s\cdot y+\bs\cdot T\cdot\by\right)\label{subeq:sObj}\\
		\text{s.t. } & y\in\{(1,0)^\top,(0,1)^\top\} \label{subeq:today}\\
		&\by_1 = 1-\by_2\in[0,1]\label{subeq:week}\\
		&p^\top \by \leq (k-k_\mathrm{ref}-p^\top y)/T\label{subeq:price}\\
		&p^\top y \leq k\label{subeq:posTmrw}\\
		&0 \leq k,\label{subeq:posTday}
		\end{align}
	\end{subequations}
\end{small}%
\endgroup
and start solving for the case where $d_1<d_2$. Since from~\eqref{subeq:today} each agent has only two possibilities (namely, arc 1 or arc 2), we first set the value of $y$ to one of these two and compute the $\by$ resulting from the combination of~\eqref{subeq:week}, \eqref{subeq:price} and~\eqref{subeq:posTmrw}. Finally, we evaluate the objective~\eqref{subeq:sObj} for both possible $y$ and resulting $\by$, and pick the $y$ minimizing it as a function of the prices $p$, discomfort $d$ and sensitivity $s$. In the following, we proceed by increasing the value of $k$.

First, a negative $k$ is infeasible due to~\eqref{subeq:posTday}.
Moreover, if $k\in[0,p_1)$, the agent cannot decrease it due to~\eqref{subeq:posTmrw}, and therefore the only possible choice is $j^\star=2$.
We can derive similar conditions for $k<k_\mathrm{ref}+p_1-T\cdot r_2$.
From condition~\eqref{subeq:week} we get
\par\nobreak\vspace{-5pt}
    \begingroup
    \allowdisplaybreaks
    \begin{small}
\begin{equation*}
    p_1\by_1 - r_2\cdot(1-\by_1) \leq \frac{k-\kref-p_j}{T},
\end{equation*}
\end{small}
\endgroup
resulting in
\par\nobreak\vspace{-5pt}     \begingroup     \allowdisplaybreaks     \begin{small}
\begin{equation}\label{eq:by1}
    \by_1\leq \frac{k-\kref-p_j+T\cdot r_2}{T\cdot(p_1+r_2)}.
\end{equation}
\end{small}%
\endgroup
Since $d_1<d_2$, \eqref{eq:by1} will hold with equality due to objective~\eqref{subeq:sObj} unless its right-hand side is strictly larger than 1.
From~\eqref{subeq:week} we have that $\by_1\geq 0$, which combined with \eqref{eq:by1} leads to
\par\nobreak\vspace{-5pt}     \begingroup     \allowdisplaybreaks     \begin{small}
\begin{equation}
    k\geq \kref-T\cdot r_2+p_j,
\end{equation}
\end{small}%
\endgroup
indicating that if $k<\kref-(T+1)\cdot r_2$ the problem is infeasible for any arc choice. On the contrary, if $k\in[\kref-(T+1)\cdot r_2,\kref-T\cdot r_2+p_1)$, the only possible choice is arc $j=2$.
In conclusion, defining $k_\mathrm{inf}=\max\{0,k_\mathrm{ref} - (T+1)\cdot r_2\}$ and $k_\mathrm{poor} = \max\{p_1,k_\mathrm{ref} + p_1 - T\cdot r_2\}$, we get infeasibility if $k<\kinf$ and $j^\star=2$ if $k\in[\kinf,\kpoor)$. Note that since traveling on arc 2 will increase $k$, an agent starting with a level of $k(0)\geq \kpoor-p_1$ will always have $k(t)\geq \kpoor-p_1$ $\forall t\geq0$.

Second, we study the case where $k\in [\kpoor,\krich]$, with $\krich = \kref+T\cdot p_1-r_2$. Here, Assumption~\ref{ass:meaningfulFlows} guarantees that the set $[k_\mathrm{poor},k_\mathrm{rich}]$ is nonempty.
Moreover, it can be seen that~\eqref{eq:by1} holds with equality as its right hand side cannot exceed 1. In fact, the condition
\par\nobreak\vspace{-5pt}     \begingroup     \allowdisplaybreaks     \begin{small}
\begin{equation*}
    \frac{k-\kref-p_j+T\cdot r_2}{T\cdot(p_1+r_2)}\leq 1 \quad \forall j
\end{equation*}
\end{small}%
\endgroup
results indeed in $p_j=-r_2$ and
\par\nobreak\vspace{-5pt}     \begingroup     \allowdisplaybreaks     \begin{small}
\begin{equation*}
    \frac{k-\kref+r_2+T\cdot r_2}{T\cdot(p_1+r_2)}\leq 1,
\end{equation*}
\end{small}%
\endgroup
yielding
\par\nobreak\vspace{-5pt}     \begingroup     \allowdisplaybreaks     \begin{small}
\begin{equation*}
    k\leq \kref + T\cdot p_1 - r_2=:\krich.
\end{equation*}
\end{small}%
\endgroup
Therefore, we get
\par\nobreak\vspace{-5pt}     \begingroup     \allowdisplaybreaks     \begin{small}
\begin{equation}\label{eq:by1eq}
    \by_1= \frac{k-\kref-p_j+T\cdot r_2}{T\cdot(p_1+r_2)},
\end{equation}
\end{small}%
\endgroup
which, combined with~\eqref{subeq:week} yields
\par\nobreak\vspace{-5pt}     \begingroup     \allowdisplaybreaks     \begin{small}
\begin{equation}\label{eq:by2eq}
    \by_2 = 1-\frac{k-\kref-p_j+T\cdot r_2}{T\cdot(p_1+r_2)} =\frac{T\cdot p_1-k+\kref+p_j}{T\cdot(p_1+r_2)}.
\end{equation}
\end{small}%
\endgroup
Next, we compute the objective from~\eqref{subeq:sObj} defined as
\par\nobreak\vspace{-5pt}     \begingroup     \allowdisplaybreaks     \begin{small}
\begin{equation}
    J(j,s,k)=s\cdot d_j+\bs\cdot T\cdot d^\top \by(j,k),
\end{equation}
\end{small} \endgroup
for $j\in\{1,2\}$ and with $\by(j,k)$ from~\eqref{eq:by1eq} and~\eqref{eq:by2eq}, and choose $j^\star$ as its minimizer.
If $j=1$, we get
\par\nobreak\vspace{-5pt}     \begingroup     \allowdisplaybreaks     \begin{small}
\begin{equation}\label{eq:J1}
\footnotesize
\begin{aligned}
    &J(1,s,k)\\
    &=s\cdot d_1+\bs\cdot T\cdot \left(d_1\cdot\frac{k-\kref-p_1+T\cdot r_2}{T\cdot(p_1+r_2)} + d_2\cdot \frac{T\cdot p_1-k+\kref+p_1}{T\cdot(p_1+r_2)}\right)\\
    &=s\cdot d_1+ \bs\cdot \frac{d_1\cdot(k-\kref-p_1+T\cdot r_2)-d_2\cdot(k-\kref-(T+1)\cdot p_1)}{p_1+r_2}.
\end{aligned}
\end{equation}
\end{small}%
\endgroup
Similarly, if $j=2$, we get
\par\nobreak\vspace{-5pt}     \begingroup     \allowdisplaybreaks     \begin{small}
\begin{equation}\label{eq:J2}
\footnotesize
\begin{aligned}
    &J(2,s,k)\\
    &=s\cdot d_2+ \bs\cdot \frac{d_1\cdot(k-\kref+r_2+T\cdot r_2)-d_2\cdot(k-\kref-T\cdot p_1+r_2)}{p_1+r_2}.
\end{aligned}
\end{equation}
\end{small}%
\endgroup
Finally, taking the difference between~\eqref{eq:J1} and~\eqref{eq:J2} yields
\par\nobreak\vspace{-5pt}     \begingroup     \allowdisplaybreaks     \begin{small}
\begin{equation}\label{eq:objdelta}
\begin{aligned}
    J(1,s,k)-J(2,s,k)&=s\cdot(d_1-d_2) -\bs\cdot \frac{(d_1-d_2)\cdot(p_1+r_2)}{p_1+r_2}\\
    &=(d_1-d_2)\cdot(s-\bs),
\end{aligned}
\end{equation}
\end{small}%
\endgroup
from which we can infer that since $d_1<d_2$, if $s>\bs$, the optimal choice is $j^\star=1$, whilst, if $s<\bs$, then $j^\star=2$.
Interestingly, the quantitative discomfort difference or the prices' values does not influence the best response strategy.

Third, we turn our attention to the case where $k\in(\krich,\kwealthy)$, with $\kwealthy=\kref +(T+1)\cdot p_1$
There we can show that condition~\eqref{eq:by1} holds with equality for $j=1$, and is inactive for $j=2$. In fact, the condition
\par\nobreak\vspace{-5pt}     \begingroup     \allowdisplaybreaks     \begin{small}
\begin{equation*}
    \frac{k-\kref-p_1+T\cdot r_2}{T\cdot(p_1+r_2)}<1<\frac{k-\kref+r_2+T\cdot r_2}{T\cdot(p_1+r_2)}
\end{equation*}
\end{small}%
\endgroup
indeed leads to
\par\nobreak\vspace{-5pt}     \begingroup     \allowdisplaybreaks     \begin{small}
\begin{equation*}
    \krich=\kref+T\cdot p_1-r_2<k<\kref + (T+1)\cdot p_1= \kwealthy.
\end{equation*}
\end{small}%
\endgroup
Therefore, choosing arc $j=1$ would still result in~\eqref{eq:by1} holding with equality, i.e., conditions \eqref{eq:by1eq} and \eqref{eq:by2eq}, yielding the objective~\eqref{eq:J1}.
Conversely, choosing arc $j=2$ would result in~\eqref{eq:by1} being inactive and hence we would have $\by_1=1$ and $\by_2=0$, yielding the objective
\par\nobreak\vspace{-5pt}     \begingroup     \allowdisplaybreaks     \begin{small}
\begin{equation}\label{eq:J2rich}
    J(2,s,k) = s\cdot d_2 + \bs\cdot T\cdot d_1.
\end{equation}
\end{small}%
\endgroup
In this case, the difference between objectives~\eqref{eq:J1} and~\eqref{eq:J2rich} results in
\par\nobreak\vspace{-5pt}\begingroup     \allowdisplaybreaks     \begin{small}
\begin{equation}\label{eq:objdeltarich}
\begin{aligned}
    J(1,s,k)-J(2,s,k)&=(d_1-d_2)\cdot\left(s-\bs\cdot\frac{\kref+(T+1)\cdot p_1-k}{p_1+r_2}\right)\\
    &=(d_1-d_2)\cdot\left(s-\bs\cdot\frac{\kwealthy-k}{p_1+r_2}\right),
\end{aligned}
\end{equation}
\end{small}%
\endgroup
from which we infer that if $s>\bs\cdot\frac{\kwealthy-k}{p_1+r_2}$ then the best response is to travel on arc $j^\star=1$, while, if the opposite is true, the optimal arc is $j^\star=2$.
Note that the sensitivity threshold is reduced from $\bs$ at $k=\krich$ to $0$ at $k=\kwealthy$.
Also in this case, the best response strategy is independent of the quantitative discomfort values.

Fourth, we consider the final case whereby $k> \kwealthy$. Since this condition results in~\eqref{eq:by1} being inactive for any arc choice, we obtain $\by_1=1$ without any constraint on the choice of $y$. Therefore, the best response strategy is trivially found to be $j^\star =1$.

Finally, combining everything together results in the best response strategy~\eqref{eq:bestResponse}.

Turning our attention to the case with $d_1=d_2$, once again we see that we get infeasibility if $k<\kinf$ and $j^\star=2$ if $k\in[\kinf,\kpoor)$.
Otherwise, since the objective does not change for any $y$ and $\by$, we get the multiple solution $j^\star\in\{1,2\}$ for any $k\geq\kpoor$.
Combining these results yields the best response strategy~\eqref{eq:bestResponseEq}.

Considering the case with $d_1>d_2$, we also get infeasibility if $k<\kinf$ and $j^\star=2$ if $k\in[\kinf,\kpoor)$.
If $k\geq\kpoor$ we can see that the objective~\eqref{subeq:sObj} is pushing both $y$ and $\by$ to $(0,1)^\top$ without being constrained by~\eqref{subeq:week}--\eqref{subeq:posTday}. This way we get $j^\star = 2$ for all $k\geq\kinf$, i.e., the best response strategy~\eqref{eq:bestResponsed1worse}.

Combining the results \eqref{eq:bestResponse}--\eqref{eq:bestResponsed1worse} concludes the proof. \qed

\section{Proof of Lemma~\ref{lem:WE2}}\label{app:proof3}
We distinguish between two cases.
First, if $x^\mathrm{BR}_1<\bx_1$, then $x^\mathrm{BR}=x^\mathrm{WE}$, since $d_1(x^\mathrm{BR}_1)<d_2(x^\mathrm{BR}_2)$.
Second, we show the existence of $x^\mathrm{WE}$ with $d_1(x^\mathrm{WE}_1)=d_2(x^\mathrm{WE}_2)$ if the number of travelers with $k<\kpoor$ is lower than $\bx_2$.
To do so, we assign the agents with $k<\kpoor$ to the second arc, since it is the only arc they can pick.
Next, we jointly fill both arcs with the remaining agents so that both arcs achieve equal discomfort, i.e., $x=\bx$. In line with~\eqref{eq:bestResponseEq}, for those agents, choosing any of the arcs would be a solution to Problem~\ref{prb:singleagent}.
Thus the resulting equilibrium is a WE with \mbox{$d_1(x^\mathrm{WE}_1)=d_2(x^\mathrm{WE}_2)$}, corresponding to the uncontrolled case.
In both cases, the WE exists and is characterized by $d_1(x^\mathrm{WE}_1)\leq d_2(x^\mathrm{WE}_2)$, concluding the proof. \qed

\section{Proof of Theorem~\ref{thm:eq}}\label{app:proof2}
First, we show that given $\Pho>0$ and any initial condition $\vP_0\geq0$ satisfying $\one^\top \vP_0=1$, the equilibrium satisfying $\Pe\geq0$ and $\one^\top \Pe=1$ is globally asymptotically stable.
Applying standard matrix theory results such as the Perron-Frobenius Theorem to column-stochastic and primitive matrices, we see that for the spectral radius of $A$ it holds $\rho(A)=1$. Furthermore, for the eigenvalues of $A$, it holds $\lambda=1> |\mu|\geq 0$ $\forall \{\lambda,\mu\}\in\rho(A)$, with $\lambda=1$ simple.
This way, we see that the time-trajectory $\vP_t$ converges to a vector spanned by $\Pe$~\cite{Bullo2018}.
Since the equilibrium probability vector $\Pe$ is spanned by the eigenvector related to the eigenvalue $\lambda=1$, it can indeed be chosen non-negative.
For the given initial condition, it holds that $\vP_t\geq 0$ and $\one^\top \vP_t=1$ for all $t\geq 0$. Combining the two, we conclude that $\Pe\geq \zero$ is globally asymptotically stable.
    
Second, we show that at the average steady-state $x^\mathrm{e}$, the ratio of people choosing the fast route $x_1^\mathrm{e}=\Pgo\cdot\one^\top A_\mathrm{rush}\Pe$ corresponds to $x^\star_1$, whilst the ratio of people choosing the slow route $x_2^\mathrm{e}=\Pgo\cdot\one^\top A_\mathrm{chill}\Pe$ to $x^\star_2$.
We know by~\eqref{eq:prices} that for the desired steady-state $x^\star$ it holds $\frac{x^\star_1}{x^\star_2}=\frac{r_2}{p_1}$.
Therefore, since $\one^\top x^\mathrm{e}=\one^\top x^\star=\Pgo$, $x^\star = x^\mathrm{e}$ holds iff
\par\nobreak\vspace{-5pt}
\begingroup
\allowdisplaybreaks
\begin{small}
\begin{equation}
\frac{x^\mathrm{e}_1}{x^\mathrm{e}_2} = \frac{r_2}{p_1}.
\end{equation}
\end{small}%
\endgroup
This means that $x_1^\mathrm{e}p_1-x_2^\mathrm{e}r_2=0$, i.e., that the total Karma level of the population will not change.
We proceed by contradiction: Suppose that at the equilibrium it holds that $x_1^\mathrm{e}p_1-x_2^\mathrm{e}r_2\neq0$. This means that the Karma distribution over the population cannot remain identical, i.e., that $A\Pe\neq \Pe$, which contradicts the fact that $\Pe$ is an equilibrium, hence proving that $x^\mathrm{e}=x^\star$.

Combining the global asymptotic stability of the equilibrium $\Pe$ with its correspondence to the system optimum $x^\star$ concludes the proof. \qed
\end{appendices}
\fi

\section*{Acknowledgments}
We thank Dr.\ Ilse New for proofreading this paper.
Furthermore, we are grateful to Dr.\ Marco Pavone for the fruitful discussions and advice.
This paper is dedicated to the loving memory of Dr.\ Kilian Schindler.

\bibliographystyle{IEEEtran}
\bibliography{../../../Bibliography/main.bib,../../../Bibliography/SML_papers.bib}

\end{document}